# When acting as a reproductive barrier for sympatric speciation, hybrid sterility can only be primary

Donald R. Forsdyke

*Department of Biomedical and Molecular Sciences, Queen's University, Kingston, ON, Canada K7L3N6*

**CONTENTS**:

**ABSTRACT**

In many animals parental gametes unite to form a zygote that develops into an adult with gonads that, in turn, produce gametes. Interruption of this germinal cycle by prezygotic or postzygotic reproductive barriers can result in two independent cycles, each with the potential to evolve into a new species. When the speciation process is complete, members of each species are fully reproductively isolated from those of the other. During speciation a primary barrier may be supported and eventually superceded by a later appearing secondary barrier. For those holding certain cases of prezygotic isolation to be primary (e.g. elephant cannot copulate with mouse), the onus is to show that they had not been preceded over evolutionary time by periods of postzygotic hybrid inviability (genically determined) or sterility (genically or chromosomally determined). Likewise, the onus is upon those holding cases of hybrid inviability to be primary (e.g. Dobzhansky-Muller epistatic incompatibilities), to show that they had not been preceded by periods, however brief, of hybrid sterility. The latter, when acting as a sympatric barrier causing reproductive isolation, can only be primary. In many cases, hybrid sterility may result from incompatibilities between parental chromosomes that attempt to pair during meiosis in the gonad of their offspring (Winge-Crowther-Bateson incompatibilities). While WCB incompatibilities have long been observed on a microscopic scale, there is growing evidence for a role of dispersed finer DNA sequence differences.

*Key words:* chromosomal speciation, germinal cycle, Haldane's rule, hierarchical barriers, meiosis, reproductive isolation

## I. INTRODUCTION

Successful sexual crossing results in offspring that are fertile and so able to continue the line. Barriers that impede this crossing of individuals within a species can facilitate the branching of that species into two species. These barrier mechanisms are broadly classifiable as chromosomal and genic. As previously noted (Forsdyke, 1999), a major advocate of a primary role for chromosomal barriers, Michael White, concluded his *Modes of Speciation* by calling for a "new synthesis" of evolutionary theory that he predicted might be manifest as a "definitive work … on speciation" appearing "about the year 2000" (White, 1978). Although far from definitive (Johannesson, 2010), there did indeed appear two major works on speciation at that time. One by a biochemist emphasised chromosomal mechanisms (Forsdyke, 2001). The other by fruit fly geneticists emphasised genic mechanisms (Coyne & Orr, 2004). Following White, the former work traced the speciation story back to Darwin, building on the studies of his research associate, George Romanes, and of the geneticist, William Bateson. As recently recapped (Coyne, 2018), the latter work built on the studies of Haldane (1922) and of Theodosius Dobzhansky (1936), whose *Genetics and the Origin of Species* had seemed to provide a reliable account of earlier studies (Dobzhansky, 1937).

However, as also recently recapped (Forsdyke, 2017a, 2018), the chromosomal viewpoint was attacked by members of the genic school (Kliman, Rogers, & Noor, 2001), and was duly defended, both by Forsdyke (2004, 2010) with further updates in successive editions of *Evolutionary Bioinformatics* (Forsdyke, 2016), and by others (Rogers *et al.,* 2018). The present review addresses the concern of the genic school that obtaining "evidence for the sequence in which [successive] reproductive barriers evolved … is usually impossible" (Coyne, 2018). It is shown here that, on theoretical grounds alone, an early role for chromosomally-based hybrid sterility in sympatric speciation may be difficult to exclude. Those emphasizing a primary role for another barrier – hybrid inviability – have tended to overlook this possibility. Furthermore, although he had coined the term "epistasis," attention is drawn to the error of attaching Bateson's name to these epistatic gene-based incompatibilities (Forsdyke, 2011; Nei, 2013). For those who appreciate eponymous acronyms a novel "WCB" terminology that correctly involves Bateson, is introduced.

Various reproductive barriers are here considered in the context of successive generational life cycles, with an emphasis on the mechanism and importance of the hybrid sterility barrier that, if operative in sympatry (Foote, 2018; Jorde *et al.,* 2018), can only be primary. When another barrier is proposed to be primary, there is an onus on the proposer to show that it had not been preceded by a period, however brief, of some degree of hybrid sterility (Forsdyke, 2017a, b; 2018).

## II. THE ADVANTAGE OF STERILITY

Since he construed natural selection as working positively when organisms survive and produce offspring, Charles Darwin found it difficult to understand how something so negative as sterility in an otherwise normal offspring (hybrid) could be favoured by natural selection (Darwin, 1862). Today, the paradox of an "inherent difficulty in genetically dissecting the phenotype that prevents its own transfer to progeny," is acknowledged (Gregorova *et al.,* 2018; Wang, Valiskova & Forejt, 2018). However, hybrid sterility in offspring can be dissected when it is regarded as a phenotypic expression, not of an offspring character, but of a character of the parents of that offspring.

While the sterility character is disadvantageous *for offspring* – for whom it would be a barrier preventing continuation of the line – sterile offspring can be the manifestation of an emergent *parental character* that, indeed, could be favoured by natural selection. The phenotype of the parents would come into play in a generational cycle immediately following their own, thus being first manifest in their offspring. As with other characters that natural selection might target, first there would have to have been internal genomic variations within parental or ancestral generations. When the character emerged, natural selection would act, either positively if the character was advantageous, or negatively if the character was disadvantageous.

While an offspring may inherit various overt characters, such as tallness or hair colour, an inherited sterility character is more subtle. When the offspring crosses with *certain members* of its species, there is a failure to produce further offspring (the potential grandchildren of its parents). However, when it crosses with *other members*, fertile offspring are produced. Likewise, those certain members with whom it fails to cross may themselves find other members with whom they may successfully cross. Thus, an individual can cross with some members of its species, but not with others. The problem is not with the individuals, but with their pairing relationships. They are reproductively compatible with some, but reproductively incompatible with others. The subtle character they manifest is *selective reproductive isolation*, a form of group selection.  How could this be favoured by natural selection?

When there exists some degree of hybrid sterility between sectors of a species, then members of that species can progressively split into two groups. *Within* each group mutual fertility exists among its members, but *between* members of different groups fertility is decreased. As proposed long ago by Romanes (1886), isolation may often have *preceded* character differentiation, not the converse (Forsdyke, 2001, pp. 47–63; Stathos & Fishman, 2014; Fuller *et al.,* 2017). When there is reproductive isolation – a term suggested to Romanes by the director of Kew Gardens (Thiselton-Dyer, 1888) – there is then the possibility of the eventual division of a species into two species that evolve different sets of characters. Members of these two species may, when viewed as a collective, meet the challenges of natural selection with a wider range of phenotypic options than the members of the one species from which they originated. In this circumstance, possession of a sterility/fertility character that facilitated speciation could be favoured by natural selection. This would be especially apparent with members of a species that had arrived at an evolutionary dead end in that they were having difficulties meeting new challenges. As suggested by Wallace in letters to Darwin early in 1868, there might be more flexibility if there were division into two species (Vortzimmer, 1970; Forsdyke, 2016, pp. 169-172).

## III. NON-MENDELIAN INHERITANCE OF THE "RESIDUE"

Aware of Romanes' work, in the first decade of the twentieth century William Bateson recognized that, unlike the transfer to offspring of more conventional characters that were distributed according to Mendel's laws, the *selective sterility* character was often distributed to *all* the offspring of a *particular* couple. Likewise, for another couple, the *selective fertility* character would be distributed to *all* offspring (Forsdyke, 2010). While Bateson sometimes found that the transmission of sterility did indeed show a Mendelian distribution, thus indicating a relationship with what we now call genes (genic sterility), this only explained hybrid sterility in a minority of cases. There was something else – a factor, apparently of a non-genic nature – that could also be manifest as hybrid sterility.

Having already coined many of the terms now familiar to geneticists – e.g. homozygote, heterozygote, epistasis – he pondered a name for that factor. The philosopher William James had written of the "unclassified residuum" that floats around the "orderly facts" of a science and "proves less easy to attend to than to ignore" (Forsdyke, 2016, p. 397). Bateson chose the name "residue" for the new factor. Furthermore, he held this likely to be fundamental to Darwin's great question: how do species originate?

## IV. WINGE-CROWTHER-BATESON INCOMPATIBILITY

A possible chromosomal basis for Bateson's residue emerged in the 1920s (Cock & Forsdyke, 2008, p. 339-377). Initiation of a speciation process requires mechanisms for achieving some degree of reproductive isolation. This decreases the probability that the process would be subverted by recombination between the genomes of diverging types. Such recombination would tend to blend (homogenize), rather than retain, the differences that were responsible for the initiation of, and/or were sustaining, the speciation process (see later). Be it of genic, or non-genic, origin, hybrid sterility is one of the three broad barriers to reproduction that can arise in sympatry among certain members of a species. An outcome of the block can be branching into two species (see later).

With the assistance of a physician (Crowther, 1922), Bateson tentatively equated non-genic hybrid sterility with an incompatibility between parental chromosomes when they pair at meiosis within the gonad of their offspring (Bateson, 1922; Cock & Forsdyke, 2008, p. 486-489). This failed homology search would be the basis of his mysterious "residue." Since a similar view had been advanced by the yeast geneticist, Öjvind Winge (1917), we can refer to this eponymously as a Winge-Crowther-Bateson (WCB) incompatibility. By contrast, the well known Dobzhansky-Muller (DM) incompatibility refers to a form of genic incompatibility involving adverse epistatic interactions between the products of parental genes (Forsdyke, 2011).

While WCB incompatibilities may involve chromosomes on a microscopically observable scale (White, 1978; King, 1993; Stathos & Fishman, 2014; Fuller *et al.,* 2017), there has been a growing focus on finer sequence changes (Forsdyke, 1996, 2001, 2016, 2017a, b, 2018; Reese & Forsdyke, 2016; see later). Furthermore, there has been growing support for the idea that meiotic "homology search and recognition can occur independently from strand invasion and genetic exchange" (Chapman *et al.,* 2017; see later). An important clue to the nature of this early homology search was evidence that a nucleic acid's base composition, rather than its actual sequence, was a major factor affecting chromosome pairing (Forsdyke, 1996, 2007; Chapman *et al.,* 2017). One of many lines of evidence came from viruses that could meet in a common cytoplasm (i.e. in 'sympatry'). Here recombination between types that had diverged from a common ancestor should be possible. If not prevented, such recombination would destroy their species individuality (i.e. they would "blend"). Retroviruses are a good example (see later).

**V. THE GERMINAL CYCLE**

For sexual organisms that are considered high on the evolutionary scale there is a clearly demarcated reproductive (germinal) cycle. While the early segregation of germline cells during animal embryogenesis is the main consideration here, degrees of segregation are also recognized in plants (Lanfear, 2018). Figure 1 shows a simplified form of the reproductive cycle in a species (A), with the cycles of the two sexes merged and their bodies, having acted as hosts to their gonads, being discarded in each generation (grey boxes). The cycling process continues within the confines of species A until one of the three fundamental barriers to cycle operation arises as a primary barrier. This primary barrier may affect many crosses until secondary, and perhaps tertiary, barriers come into play. There is a progressively irreversible divergence into new lines with independent cycles that are seen today as species B and C.

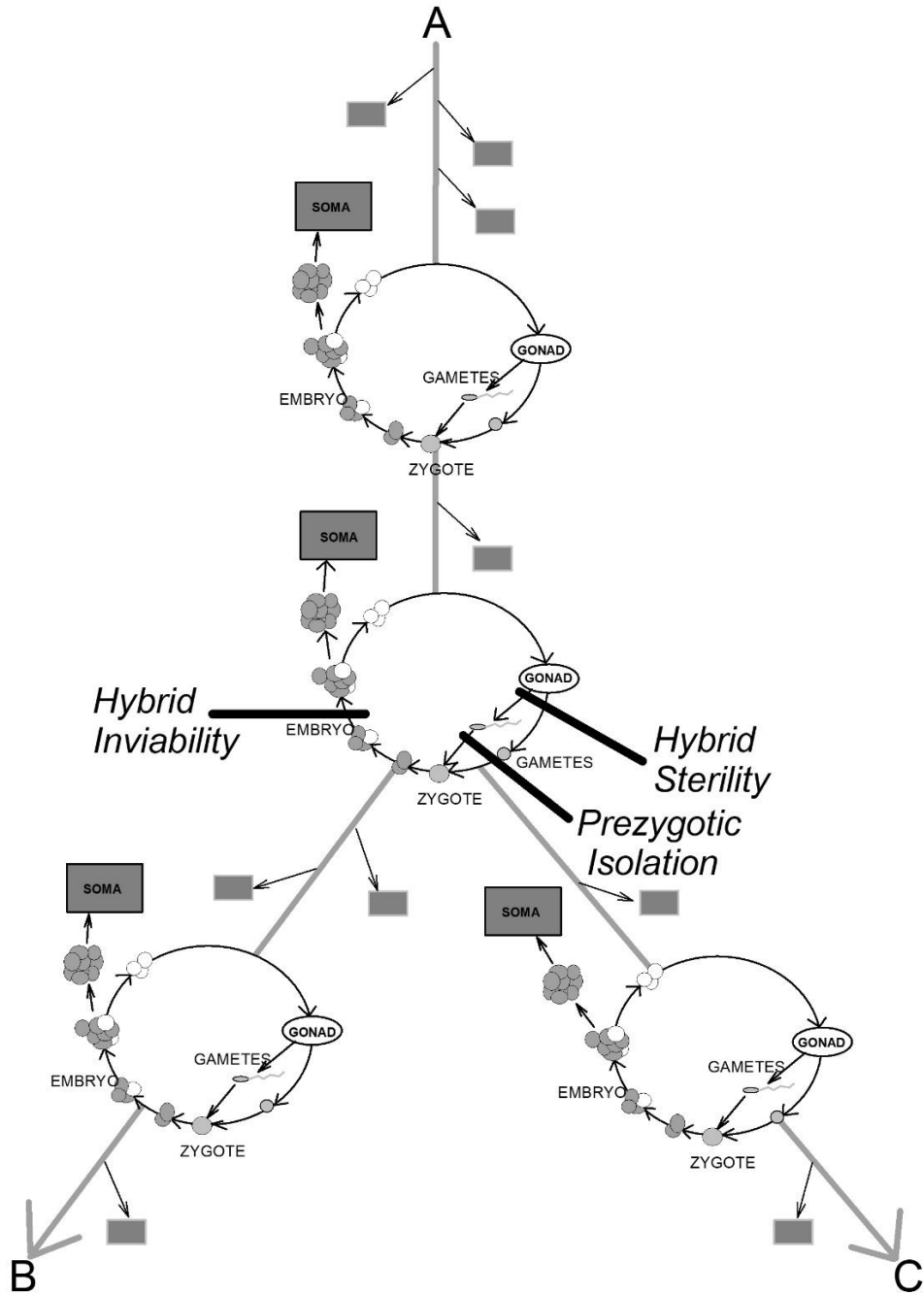

**Fig. 1.** Branching of species A into new species (B and C). The mortal soma (grey boxes) provides support for the gonad, but is discarded in each generation (small grey boxes symbolizing numerous discardments). The germinal cycle (top) operates continuously to facilitate within-species sexual crossing until there appears one of three barriers (bars). These have the potential to slow or stop production of fertile offspring by preventing either gamete transmission (prezygotic isolation), or zygote development (hybrid inviability), or gamete formation (hybrid sterility). When speciation is complete (bottom), two independent germinal cycles maintain the continuity of species B and C through the generations. To simplify, male and females here share a common cycle.

When viewed abstractly, cycles operate continuously and there may be no obvious start or end-points. A blockage at any one of three main points (1, 2 or 3) either halts or slows a cycle (Fig. 2a). When halted (i.e. no species member produces offspring), subsequent downstream points of potential blockage cannot contribute to cycle arrest. When slowed (i.e. some species members are less susceptible and their offspring people the next generation), these subsequent points of potential blockage may contribute to, and eventually replace, the point of initial blockage that occurred in an earlier generational cycle. They become *secondary* blockage points that, likewise, may eventually be replaced by *tertiary* blockage points in later generational cycles.

The actual reproductive cycle differs from the abstract and is shown in Fig. 2b with separate cycles for the two sexes (α and α that have closely related DNA sequences). A cycle operates in an individual of one sex and interacts, through its generation of gametes, with the reproductive cycle that operates in an individual of another sex. Both cycles are naturally interrupted at one point (1) when gamete transmission occurs, and a new cycle emerges in the form of an individual (α) with some of the characteristics of the parenting individuals (upper part of Fig. 2b). This process can recur, so establishing a lineage of α organisms.

With each crossing there is some assortment of parental characters to constitute a new offspring and, while there may be no loss of character-forming *potential*, in classical Mendelian terms there is no blending of characters. This is seen most obviously in the inheritance of sex. Although her father was male, a daughter does not inherit her father's sex. Although his mother was female, a son does not inherit his mother's sex. Multiple genes are involved in sex determination, but various features of the sex chromosomes tend to impede their recombinational re-assortment, so the genes act as a unit and there is no blending.

However, when multiple genes contribute to a character (such as height in humans) then, as Mendel recognized, blending is usual (Forsdyke, 2016, pp. 145-146). In such circumstances, a tendency for divergence into different forms (say tall and short in humans) will be countered by the blending that occurs (e.g. when a tall person crosses with a short person). For different lines to emerge, tall must cross with tall and short must cross with short. To achieve such *selective* crossing, some externally or internally imposed form of reproductive isolation (reproductive selection) must occur. There must be barriers.

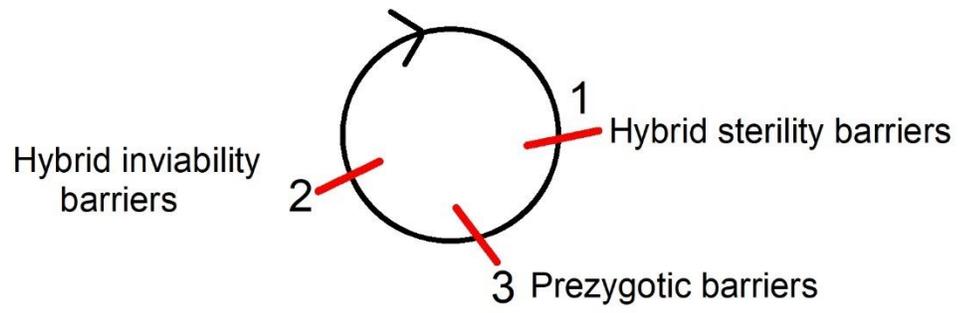

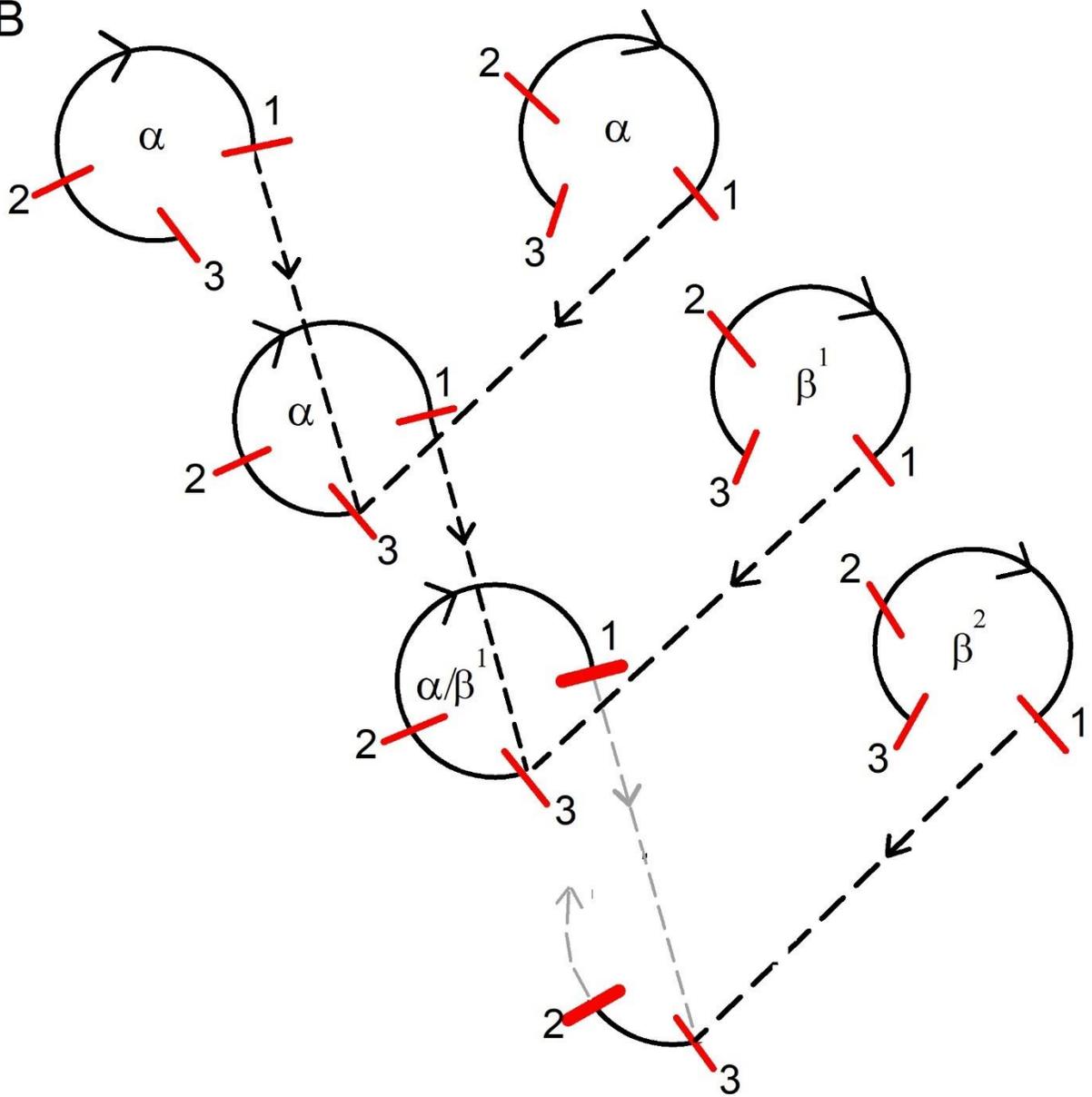

**Fig. 2.** Roles of hierarchically related barriers in interruption of germinal cycles. Since the germinal cycle is recursive, *any point*, be it before or after union of gametes to form a zygote, can mediate primary arrest of the cycle (points 1, 2 and 3; red lines in Figure 2A). The arrow indicates clockwise progression, but numbering is anticlockwise. When 3 is primary the cycle arrests and downstream events (2 and 1) cannot occur in that individual. When 2 is primary, then downstream event (1) cannot occur in that individual. However, when 1 is primary it cannot affect 3 and 2 *in the same individual* because sexual collaboration with a partner (α with α) generates a new cycle (upper part of Fig. 2B). Thus, it is normal for the cycle to interrupt when a hybrid sterility barrier (1) is absent. If prezygotic (transmission) barriers (3) are absent, gametes from pairing partners can meet (dashed black lines) as a new individual (another α with a new cycle). If hybrid inviability barriers (2) are absent in that new individual, then gonadal gametogenesis can follow, provided hybrid sterility barriers (1) are still absent. The cycle can then resume in another α individual, and so on (not shown). However, collaboration with a partner with a slightly divergent DNA sequence (α with $β^1$) may generate a new cycle (α/$β^1$) that arrests at the hybrid sterility barrier (middle thick red line). In this circumstance, to continue their respective lines, α/$β^1$ must meet another α/$β^1$ (not shown), and $β^1$ must meet another $β^1$ (not shown). Should there be further DNA divergence ($β^1$ to $β^2$) then a collaboration with an α (not shown) or with the few α/$β^1$ offspring who chance to evade their cycle's hybrid sterility barrier (shown as dashed grey line), could be stopped by the hybrid inviability barrier (2; bottom thick red line). This new cycle's hybrid sterility barrier 1 is then excluded. To continue their lines, α/$β^1$ must find another α/$β^1$ (not shown), and $β^2$ must find another $β^2$ (not shown). With further DNA divergence ($β^2$ to $β^3$; not shown), the transmission barrier (3) can come into play, so excluding 2 and 1.

We are here concerned with natural, internal, non-geographic, barriers to crossing that can be hierarchically related (primary, secondary, tertiary). Among these intrinsic barriers, hybrid sterility is special in that, when manifest, it is primary. Within one organism, the sterility barrier cannot be replaced by another barrier, yet, within a subsequent organism in the line, another barrier (secondary) may arise to pre-empt the emergence of hybrid sterility in that organism. Likewise, within a subsequent organism, a tertiary barrier may arise to pre-empt the re-emergence of the secondary barrier within the line (Fig. 2b lower).

Defective gamete transmission always trumps inviability, since without transmission there can be no zygote to develop into an embryo. And if there is no embryo there can be no adult for gametogenesis. Even if transmission required far fewer genes than development (thus less opportunities for barrier-creating mutations), over long evolutionary time scales mutations in transmission-related genes would be bound to occur. Accordingly, when genic differences underlie cycle interruption, those defects are most likely to *finalize* among transmission-related genes. As far as reproductive isolation is concerned, any pre-existing barrier-creating differences in genes acting post-zygotically then become irrelevant.

Indeed, many modern species have ended up controlled by barrier 3 where multiple genes have failed to co-adapt (e.g. elephant cannot copulate with a mouse). However, in others 2 remains the identified barrier. Neither identification excludes the possibility of an earlier role for 1 as a fundamental barrier acting at the time of an initial divergence into two species. Although generally unlikely, even barrier 1 has sometimes persisted (see later). When there are successive barriers, the temporal sequence may sometimes be 1, 2, 3. Once pre-empted, early acting barriers are liberated from their reproductive isolating role.

Thus, regarding the speciation process, there is a "multi-dimensional … continuum" (Mérot *et al.,* 2017). Although an early member of a hierarchical sequence may sometimes be forestalled (pre-empted) by a later, there should be evidence that the early member had not acted before concluding that it was not primary. A clear determination that a hybrid sterility barrier was absent at initiation may prove difficult since, once liberated from its reproductive isolation role, it may respond to other pressures (Forsdyke, 2001, p. 22). However, new approaches to the sequencing of "natural DNA archives" (Olajos *et al.,* 2017), may now permit a finer analysis of sequence changes during speciation.

## VI. SEX CHROMOSOMES AND HALDANE'S RULE

A primary role for hybrid sterility was suggested from the generalization that, prior to the development of full sterility among offspring (F1 hybrids), the sterility may be partial in that only one sex (male in mice) is affected (Haldane, 1922). Furthermore, experimental exchanges ("introgressions") of individual chromosomes between "closely related mouse subspecies" have shown that "heterospecific autosomal pairs in sterile hybrids are more prone to asynapsis than the homospecific pairs in which both homologs came from the same species" (Bhattacharyya *et al.,* 2013). Even among heterospecific pairs, some chromosomes appear more disposed to asynapsis than others. Thus, Bhattacharyya *et al.* (2014) note that "the number of unsynapsed autosomes per cell varies, indicating the same type of cis-acting mechanism operates on individual autosomes." Their observations are attributed, not to some mobile ("trans-acting")

12genic factor with the ability to single out individual chromosome pairs, but to "their fast evolving non-genic divergence," which could have affected some chromosome pairs more than others (Bhattacharyya *et al.,* 2013).

Being *already* disparate, the sex chromosomes in male mice (XY) regularly fail to pair along most of their lengths. Relative to females with homologous sex chromosomes (XX), this gives males (the "heterogametic" sex) a head-start along the path to full sterility (Haldane's rule for hybrid sterility; Forsdyke, 2000). While conceding an important role of genes in this process (*prdm9* and a hemizygous gene on the male X chromosome), Bhattacharyya *et al.* (2013) conclude that "variation in pairing failure is under genic control," but the sterility itself "is chromosomal, caused by heterospecific pairing incompatibilities." They deem this supportive of similar suggestions regarding sterility in both fruit fly (Naviera & Maside, 1998; Moehring, 2011), and yeast ("simple sequence divergence acted upon by the mismatch repair system;" Louis, 2009; Rogers *et al.,* 2018).

Whether sex chromosomes are the same size (homomorphic) or of different size (heteromorphic), these considerations also apply to plants with independent sexes (Ironside & Filatov, 2005). Indeed, Delph and Demuth (2016), claiming to have "the best explanation for male rarity in some *Silene* hybrids," consider that "although the original chromosomal mechanism … largely fell out of favor, recent work has argued for its importance on theoretical grounds." Furthermore, demonstrating the progressive involvement of autosomes *after* a primary involvement of sex chromosomes, Hu and Filatov (2016) find an "increased species divergence and reduced gene flow on the *Silene* X-chromosome," but gene flow involving autosomal loci is still "sufficient to homogenize the gene pools of the two species."

## VII. THE PAIRING MECHANISM

Disparities in the DNA sequences that parents have contributed to their offspring can suffice to impair the chromosome pairing needed for error-correction (Bernstein, Bernstein, & Michod, 2017; Brandeis, 2018). When the disparity within their offspring's gonad is not correctable, then only within the bounds of an emerging new species will each parent be likely to find a non-disparate partner.

For successful homologous recombination two DNA duplexes must pair and exchange segments. This requires both recognition of some degree of sequence similarity and strand breakage. The temporal order of these events is contentious. A popular model postulates cutting to produce a single strand that *then* seeks a pairing partner in the homologous duplex (Szostak *et al.,* 1983). However, a growing view, summarized by Zickler and Kleckner (2015), suggests otherwise:

> A prominent, but still mysterious, feature of chromosome biology is the ability of homologous chromosomes, or chromosomal regions, to specifically recognize and pair with one another in the apparent absence of DNA lesions (DSBs) or recombination. … Recombination-independent pairing … plays prominent roles for premeiotic and meiotic programs, where it is defined as pairing that occurs before and/or without … DSBs.

A "cut first" model implies a *localized commitment* prior to pairing. A "pair first" model should more reliably afford reversible genome-wide homology *testing*, without commitment (McGavin, 1977; Wilson, 1979; Boeteng *et al.,* 2013). An initial alignment through "kissing" interactions between the loops of extruded DNA stem-loop structures requires only loop-loop base-pair complementation (Forsdyke, 2007). Thus, initial *breakage-independent* homology recognition might not lead to strand breakage and segment exchange. This would require that dispersed loop homologies be interspersed with stem sequences that were also homologous.

Indeed, from studies of homology-directed DNA changes in fungi (repeat-induced point mutation; RIP), Gladyshev and Kleckner (2014) provide "a new perspective … for … the breakage-independent recognition of homology that underlies RIP and, potentially, other processes where sequence-specific pairing of intact chromosomes is involved." Thus, "the nucleotide composition of participating DNA molecules is identified as an important factor," and "homology recognition is modulated by the underlying sequence" (Gladyshev & Kleckner, 2016). Accordingly, "sequence information can be compared directly between double-stranded DNA molecules during RIP," and there is the potential for application to "other processes where homologous pairing of intact DNA molecules is observed." This view is supported by later studies (Gladyshev & Kleckner, 2017; Chapman *et al.,* 2017).

However, the "pair first" and "cut first" views on the process by which genomes of paternal and maternal origin exchange information in the gonad of their child, should not necessarily be mutually exclusive. A limited number of pair-first sites ("buttons") might suffice (i.e. provide anchor points) to assist the close apposition of homologous chromosomes. Once this alignment was achieved, the homology-search task of single-strands liberated by a "cut first" mechanism should be easier (i.e. they could then "zipper" up; Viets *et al.,* 2018).

Ideal evidence on the mechanisms through which sympatric homology searches might fail would involve species where primary sympatric hybrid sterility barriers had not been superseded by later acting barriers. As indicated earlier, retroviruses can help here.

## VIII. BASE COMPOSITION OF RETROVIRUSES

It is increasingly recognized that "viruses, like the cellular organisms they infect, assort into reproductively isolated groups and can be organized into biological species" (Bobay & Ochman, 2018). When we compare two viral species that have a *common* host cell (i.e. "sympatric"), with two viral species that, even within a common host, do not share a common cell, we would expect to observe a fundamental difference related to their reproductive isolation mechanisms. If that difference is found to apply to other viral pairs that occupy a common host cell, then a fundamental isolation mechanism may have been identified. Retroviruses provided an example (Forsdyke, 1996).

An important measure of base composition is the proportion of the individual bases G and C, among the four bases (A, C, G, T). This is expressed as GC%. The AIDS virus, HIV1, has a low GC% value. Thus, in its DNA form it is AT-rich. This mainly reflects an increase in A, which is a purine (R), so the "R-loading" index is very high (Fig. 3). In contrast, the T cell leukaemia virus, HTLV1, has a high GC% value. Its GC-richness largely reflects an increase in C, which is not a purine, so the index of R-loading is very low.

The host of both these viruses is the T-lymphocyte. Assuming their evolution from a common ancestor and a continuity in their need to frequent a common host cell, then initial small differences in base composition could, either directly or indirectly, have played a role in preventing recombination between them within that host cell (analogous to a hybrid sterility barrier). This weak primary barrier to their recombination seems not to have been followed by the emergence of an effective secondary barrier. Their gene products have not adversely interacted (no 'hybrid inviability'). Furthermore, although there are mechanisms to prevent

'superinfection' of an already infected cell (Forsdyke, 2016, p184-186), their co-entry into a host cell was not absolutely denied (no 'prezygotic isolation'). Rather, the primary barrier would have been progressively strengthened. This is a likely explanation for the extensive base-composition difference that we see between today's co-infecting viruses (insect viruses, herpes viruses, retroviruses; Forsdyke, 1996). It is tempting to extend this explanation to the wide difference in GC% of certain protozoal parasites that undergo meiosis within a common host (Lee, Mortimer, & Forsdyke, 2004).

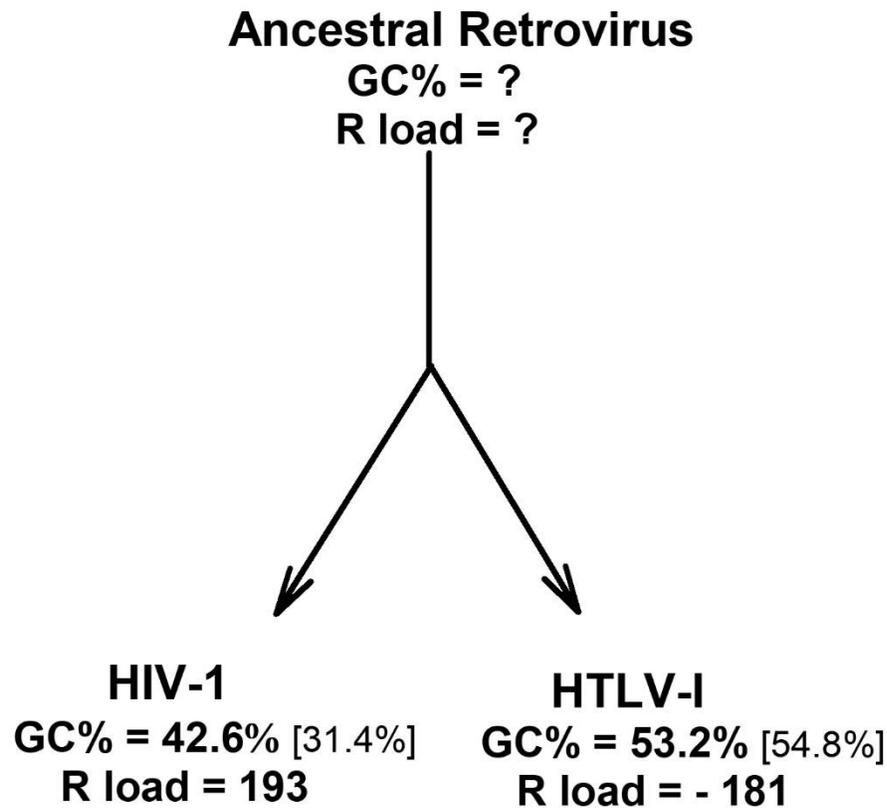

**Fig. 3.** Extreme differences in base compositions of two modern viruses that are presumed to have arisen from a common ancestor and are still capable of co-occupying a cell where mutual recombination could threaten their integrities. Base composition values (GC%) are from Bronson and Anderson (1994). Purine-loading (R-loading) indices are from Cristillo *et al.* (2001). Base compositions of the more flexible third positions of codons are shown in parentheses. HIV1 preservation correlates with greater AT-richness. HTLVI preservation correlates with greater GC-richness. These base compositions correlate with specific oligonucleotide frequencies. A mismatch between oligonucleotides would impede the DNA 'kissing' interactions that precede homologous recombination.

Although there may be phenotypic differences, if these do not emerge as effective secondary barriers, then primary barriers remain operative, not yet liberated from their reproductive isolation role. When there are no obvious phenotypic differences, the speciation process, deemed "cryptic," may continue to depend on these primary barriers. Thus, it would be predicted that, if based on general sequence differences, these would be conserved and strengthened. On the other hand, the pressure to conserve would decrease if genic differences emerged that were able to constitute secondary barriers. Indeed, when comparing sympatric populations, genomic divergence is found to be higher between cryptic than between non-cryptic populations (Jorde et al. 2018).

General base composition differences are displayed not only with individual bases, but also with groups of bases (oligonucleotides). Thus, in HIV1 we find higher frequencies of AT-rich oligonucleotides (e.g. AAT, AAA, TAA); in HTLV1 there are higher frequencies of GC-rich oligonucleotides (e.g. CCG, CCC, GCC). Species differences in these oligonucleotide frequencies, rather than in the individual bases (Brbić *et al.,* 2015), should be critical in preventing recombination and thus maintaining the integrity of virus species (Forsdyke, 2016, p.175-192; Forsdyke, 2017b). Thus, these oligonucleotides seem the best molecular candidates for the Bateson's "residue" role. Differences between these "buttons" would keep species apart.

Criteria employed for species classification do not necessarily relate to how those species initially formed. However, methods based on genotype rather than phenotype might fulfil this role. Hence it may not be incidental that oligonucleotide (tetramer) frequency has emerged as a metagenomic species classifier of high reliability (Wang, Herbster, & Mian, 2018; Forsdyke, 2019).

## IX. CONCLUSIONS

(1) Any barrier in the germinal cycle can be primary in its potential to lead to divergence into two species. An exception is when reproductive isolation is first imposed by geographical separation. In this case the external allopatric barrier is primary.

(2) In sympatry, the internal hybrid sterility barrier *when operative* can only be primary. It should be formally excluded before concluding that another barrier is primary. In that hybrid sterility may have a chromosomal, rather than genic, basis, this requires that WCB incompatibilities be excluded.

(3) Whether or not the controversy between chromosomal and genic schools persists, it seems there is now agreement that **"studying the old and classic work of evolutionary genetics remains eminently worthwhile: many puzzling phenomena and problems ripe for study lie buried in that literature"** (Coyne, 2018).

## X. ACKNOWLEDGEMENTS